# Spalled barium titanate single crystal thin films for functional device applications

*Prachi Thureja,[1] Andrew W. Nyholm,[1] Martin Thomaschewski,[1] Phillip R. Jahelka,[1] Julie Belleville,[1] and Harry A. Atwater[1,\*]*

[1]Thomas J. Watson Laboratories of Applied Physics, California Institute of Technology, Pasadena, California 91125, USA

\*Corresponding author: haa@caltech.edu

## Abstract

We report a scalable approach for fabricating single-crystal barium titanate (BTO) thin films through spalling from bulk substrates. Conventional thin film growth techniques often face challenges in achieving high-quality single crystal microstructure over large areas, resulting in reduced performance in functional devices. In contrast, spalling – *i.e.*, performing stress-induced exfoliation of bulk single crystals – enables the separation of single crystal thin films with controllable thicknesses ranging from 100 nm to 15 μm and lateral dimensions up to several millimeters. Electro-optic characterization of the spalled films yields a Pockels coefficient of $r_{33}$ = 55 pm/V in multi-domain regions and 160 pm/V in single-domain regions, leading to projections up to 1980 pm/V for $r_{42}$ under conditions of unclamped excitation. Our results indicate that spalled BTO single-crystal thin films preserve bulk electro-optic properties and exceed the performance of commercially available thin-film lithium niobate, making them suitable for integration in advanced photonic and optoelectronic devices.

Rapid advances in information technology necessitate functional devices that can meet the growing demands of modern data processing and communication systems. These devices must operate at ultrafast speeds, consume minimal power, and integrate seamlessly into compact architectures. Beyond conventional electronics, photonic platforms hold the promise to address these requirements, offering high-speed operation, low energy dissipation, and high accuracy.[1–4] Achieving such capabilities requires integrating optical materials with strong electro-optic responses in thin-film form, essential for low-power photonic devices.

Silicon on insulator (SOI) technology is widely adopted in photonic circuits due to its compatibility with CMOS processes.[5–7] However, its optical tunability is limited, relying on thermo-optic effects [8,9] or material doping,[10] which typically incur latency or loss. Transition metal dichalcogenide (TMDC) monolayers offer gate tunability, strong light-matter interactions, tunable bandgaps, and short response times.[11–13] Despite these advantages, TMDC-based devices often exhibit narrow operational bandwidths, limiting their versatility in broadband applications, and require large interaction lengths due to their atomically thin active medium. Lithium niobate ($LiNbO_3$, LN), long regarded as the benchmark for bulk electro-optic applications, has gained renewed interest in its thin-film form. Thin-film LN enables enhanced performance in nanophotonic devices through an increased integration density.[14–16] However, its modest electro-optic coefficient of $r_{33}$ = 30 pm/V[17] necessitates extended light-matter interaction lengths (often exceeding 100 µm[18]), thus posing limitations for compact, low-power designs.

These challenges underscore the need for advanced materials with superior optical properties. Barium titanate ($BaTiO_3$, BTO) is a perovskite oxide with exceptional dielectric, ferroelectric, and electro-optic properties.[19–22] In its non-centrosymmetric tetragonal phase ($a = b \neq c$-axis), stable between 5°C and 120°C, BTO exhibits a spontaneous polarization that gives rise to a pronounced electro-optic response. The Pockels coefficients of bulk BTO ($r_{42}$ = 1300 ± 10 pm/V, $r_{33}$ = 105 ± 10 pm/V, $r_{13}$ = 10.2 ± 0.6 pm/V)[4] far exceed those of LN. BTO further supports ultrafast modulation capabilities,[23,24] a broad transparency window (0.4 – 7 µm), and high refractive index values,[4,25] making it a strong candidate for many photonic and optoelectronic devices.

| Synthesis method | Refractive index ($\lambda$ = 1550 nm) | Electro-optic coefficient $r_{42}$ (pm/V) | Modulation frequency |
|---|---|---|---|
| Bulk (top-seeded solution growth) | 2.286 [26] | 1300 [4] | < 0.1 GHz[27] |
| MBE[28] | 2.286 | 923 | 65 GHz |
| PLD[29] | 2.27 | 390 | 70 GHz |
| RF sputtering[30] | 2.278 | 89 | 419 kHz |
| CVD[31] | 2.13 | 4.5 | 17.3 kHz |
| CSD (sol-gel)[32] | (1.94 at $\lambda$ = 633 nm) | ($r_{eff}$ = 27 pm/V)* | 5 MHz |

**Table 1. Refractive index and electro-optic coefficient $r_{42}$ for BTO synthesized using various techniques.** The reported refractive index values represent the average of the ordinary and extraordinary indices. The electro-optic coefficients correspond to measurements performed at the modulation frequencies specified in the adjacent column. *For sol-gel BTO, an effective electro-optic coefficient is reported due to the polycrystalline nature of the material.

To harness these properties in practical devices, however, BTO must be integrated in thin-film form. Conventional deposition techniques often result in polycrystalline or amorphous films with reduced electro-optic performance compared to their single-crystal counterparts. The extent of this reduction is highly dependent on the growth method.[33,34] Table 1 provides an overview of the $r_{42}$ coefficient reported for BTO prepared using various growth techniques. Currently, molecular beam epitaxy has yielded the highest reported electro-optic coefficient of $r_{42}$ = 923 pm/V,[28] thought the method remains slow and expensive, limiting its scalability.

To overcome the challenges associated with thin-film fabrication, we propose spalling to produce large-area, single-crystalline BTO thin films from bulk substrates. This low-cost, scalable technique offers a path toward preserving bulk electro-optic coefficients, comparable to those obtained through epitaxial growth. Spalling, initially developed for III-V semiconductors, is a kerfless method for separating layers from their host substrates.[35] In this process, a metal stressor layer is deposited on a bulk crystalline substrate, introducing compressive stress within the substrate. When the residual stress surpasses a critical threshold, a crack can be initiated with an external pulling force, such as exfoliation with adhesive tape.

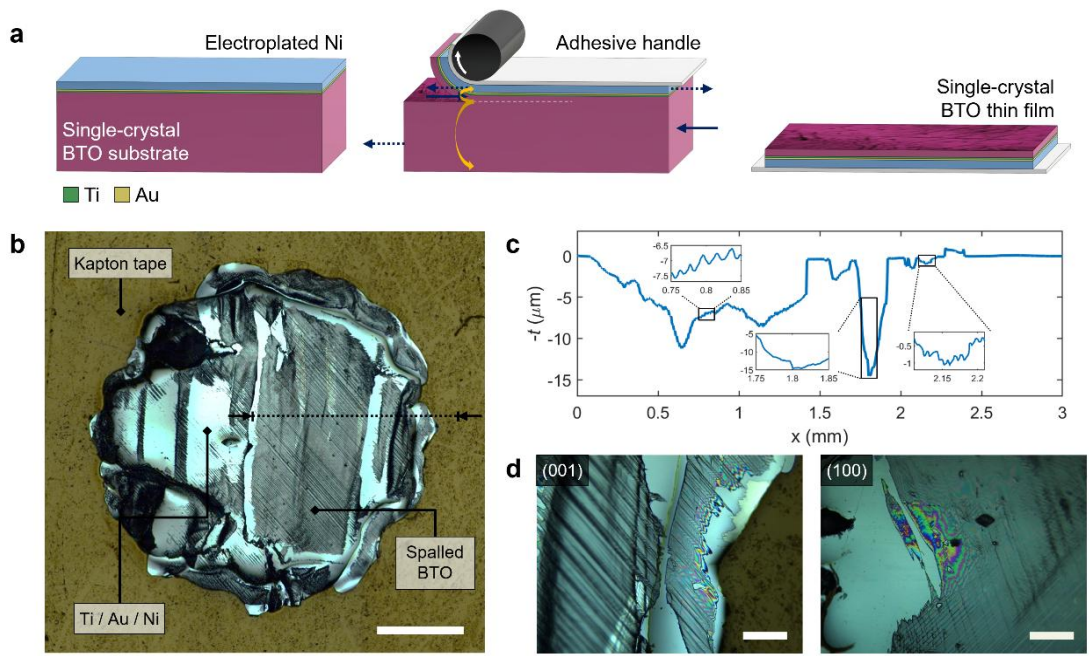

**Figure 1. Spalling barium titanate thin films.** (a) Schematic representation of procedure for spalling BTO. Left: A seed layer of Ti (green) and Au (yellow) is deposited on a bulk single-crystal BTO substrate. Ni (blue) is electroplated on top of the seed layer to induce compressive stress in the substrate. Middle: An adhesive handle (light grey) is stuck onto the substrate and can be peeled off with a roller (black). A force balance between the tensile (dotted, black) and compressive forces (solid, black) as well as the bending moment (orange) dictates the spall depth (white dashed). Right: The spalled single-crystalline BTO thin film is attached to the metal back layer and an adhesive handle. (b) Optical microscope image of spalled (001)-oriented BTO film. Scale bar: 1 mm. (c) Profilometer scan along black dotted line in (b), performed on the spalled substrate. The y-axis indicates the negative of the spall depth, -t, with zero corresponding to the level of the unspalled substrate. Insets show magnified height profiles corresponding to the square boxes, highlighting variations in spall depth and surface roughness. (d) Optical microscope images with spalled regions indicating thin film interference effects for (001)- and (100)-oriented films on the left and right, respectively. Scale bars: 100 μm.

Figure 1a illustrates a schematic of our approach, which begins with depositing a 30 nm Ti and 30 nm gold (Au) seed layer on a single-crystal BTO substrate. This conductive base layer enables electroplating on the otherwise insulating BTO. We then use a heated nickel chloride solution to electroplate 1-4 μm of Ni. The Ni thickness is primarily controlled by varying the plating time while other conditions remain fixed. The thickness of the resulting spalled film, also referred to as spall depth $t$, is determined by a force balance between the tensile force in the electroplated Ni, the induced compressive force in the BTO, and the opening mode fracture introduced through an external pull force.[35,36] In our setup, this pull force is applied using a roller-mounted tape, which enables exfoliation of continuous, single-crystalline BTO thin films. X-ray diffraction (XRD) confirms their crystallinity (Supplementary Information Part 1).

The lateral dimensions of spalled BTO films currently range up to a few millimeters, as dictated by the 4.5 mm of the plating aperture. However, prior studies on III-V semiconductors show that wafer-scale spalling is feasible.[37,38] Additionally, the same bulk BTO substrates can be reused through vibration polishing or alternative chemical-mechanical polishing methods (see Supplementary Information Part 2), further improving the cost efficiency of this approach.

Figure 1b shows a spalled (001)-oriented, single-crystalline BTO thin film on top of the Ti/Au seed layer, followed by Ni and Kapton tape (from top to bottom). Optical microscopy reveals a large, continuous BTO area with lateral dimensions of approximately 1.5 x 3 mm$^2$, alongside smaller spalled regions. Discontinuities are attributed to nonuniform Ni plating and inconsistencies in the manual pull force. While Kapton tape was used here, alternatives like polyvinyl alcohol (PVA) or heat-release tape could enable easier transfer and will be explored in the subsequent discussion. A profilometer scan of the spalled substrate (Fig. 1c) along the dotted black line in Fig. 1b shows a spall depth ranging from 15 μm (middle inset) to below 1 μm (right inset). Surface roughness varies from hundreds of nanometers in thicker regions (left inset) to tens of nanometers in thinner films (right inset). Sub-micron-thick regions additionally exhibit bright color variations under white light, indicative of thin-film interference effects. These observations are consistent across both (001) and (100) oriented films spalled from single-crystal bulk substrates, as shown in Fig. 1d.

The brightly colored regions at the edges of spalled films highlight the potential for producing larger-area films with sub-micron thickness suitable for photonic integration. Achieving such films requires precise control over the Ni thickness and the corresponding stress induced in the substrate. To explore this further, we electroplated two partially overlapping Ni layers (approximately 1.35 μm and 1 μm thick, Fig. 2a inset) on a $c$-axis BTO substrate, producing three regions with distinct stress profiles (from left to right: intermediate, high, and low). Spalling all three simultaneously minimized artifacts from variation in the manual pull force.

Figure 2a shows an optical microscope image of the spall performed using PVA tape. The highest stress region (blue box) produced the largest spalled area, spanning millimeter-scale lateral dimensions with 5-7 μm thickness. The intermediate (orange box) and low-stress states (green box) yielded continuous films extending several hundred micrometers, with thicknesses between 0.4-2.5 μm and 0.1-1.8 μm, respectively. Notably, the lowest stress state resulted in the thinnest and smoothest films, characterized by brightly colored areas exceeding 100 x 100 μm$^2$ (Fig. 2b). To quantify surface roughness, we performed atomic force microscopy (AFM) across the three regions in Fig. 2b. As shown in Fig. 2c, the root mean square (RMS) roughness $r$ decreases with film thickness. Specifically, $r$ corresponds to 125 nm in the high-stress region, 66 nm in the intermediate-stress region, and 27 nm in the low-stress region. The RMS roughness was evaluated over the areas marked by white dashed boxes in Fig. 2c. A similar analysis for spalls obtained from (100)-oriented bulk crystals is provided in Supplementary Information Part 3.

Additional AFM scans across varying film thicknesses allowed us to establish a relationship between roughness $r$ and thickness $t$ (Fig. 2d). Films spalled from both (001)- and (100)-oriented single-crystal substrates followed similar trends, with roughness scaling as $r(t) \sim C \cdot t^{0.5}$. The prefactor $C$ equals 55.8 and 74.1, respectively, for (001) and (100) orientation, while the exponents are 0.45 and 0.40. Higher roughness in (100) films likely arises from a more inhomogeneous domain structure in the initial bulk substrate (see Supplementary Information Part 4).

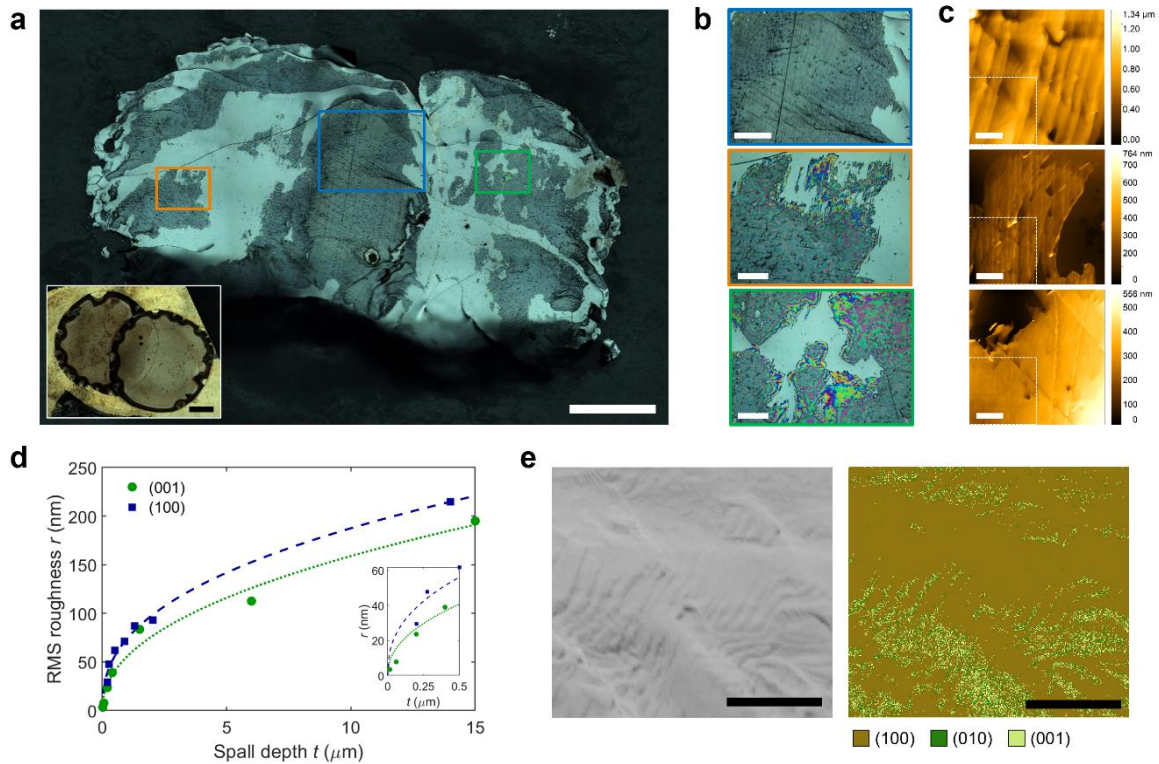

**Figure 2. Physical characterization of spalled BTO films.** (a) Optical microscope image of spalled (001) BTO thin film. The orange, blue, and green boxes correspond to regions obtained through spalling from intermediate, thick, and thin Ni stressor layers, respectively. The inset illustrates the two overlapping circular electroplated Ni layers on top of a Ti/Au seed layer on the BTO substrate. The thickness of the electroplated Ni from left to right corresponds to 1.35 μm (intermediate), 2.4 μm (thick), and 1.1 μm (thin). Scale bars: 1 mm. (b) Magnified optical microscope images of the area in the blue, orange, and green boxes in (a). Scale bars (from top to bottom): 250 μm, 100 μm, 100 μm. (c) AFM images of the spalled (001) thin films from the areas shown in (b). The dashed white boxes indicate regions over which the RMS roughness was evaluated. Scale bars: 10 μm. (d) RMS roughness $r$ of spalled (001) and (100) BTO films as function of the spall depth $t$. The inset highlights the roughness for spall depths below 500 nm. (e) Band contrast image (left) and corresponding crystal orientation map (right) obtained through electron backscatter diffraction. Scale bars: 20 μm.

Previous studies have reported a linear correlation between spall depth and Ni stressor thickness when the stressor is significantly thinner than the substrate.[35,39] This stems from greater compressive strain energy introduced into the substrate with increasing stressor thicknesses.[35,40,41] In ferroelectric materials such as BTO, such compressive strain can induce domain switching.[42,43] To investigate this further, we performed electron back scatter diffraction (EBSD) on a film spalled from a (100)-oriented bulk crystal. Figure 2e shows the band contrast image of the scanned area, reflecting the film's topography, alongside a crystal orientation map of a 60 x 60 μm² area. A large single-domain area (approximately 40 x 15 μm²) with (100) orientation is visible in the upper half of the scanned region. Variations in domain orientation can be correlated with topographical features.

These observations, combined with prior studies, suggest a strong correlation between Ni thickness, and BTO film thickness and roughness. However, further studies are necessary to map the strain depth profile in the BTO substrate and clarify whether domain switching is a static process that arises during electroplating of the stressor layer, or a dynamic process occurring as strain energy is released during exfoliation. Understanding this distinction will be key to further tuning surface and domain properties of spalled BTO films. Finally, we note that the initial quality and domain configuration of the bulk substrate significantly influence spalled film roughness. Using high-quality, ideally single-domain bulk substrates is therefore essential for producing smooth spalled thin films of ferroelectric materials.

For functional device fabrication, spalled films must be transferred from the exfoliation tape onto a solid substrate. We use water-soluble PVA tape for exfoliation, allowing easy removal. The spalled film, adhered to the PVA tape, is placed on a glass slide and spin-coated with polypropylene carbonate (PPC) to ensure uniform adhesion of the entire heterostructure to a polydimethylsiloxane (PDMS) stamp (Fig. 3a, d). The desired region is cut out with a blade and picked up using the PDMS stamp, which is subsequently immersed in water to dissolve the PVA tape.

For transfer, we use a UV-curable resin that is semi-resistant to solvents. The PDMS stamp is aligned over a drop of resin on the target substrate, with the Ni layer facing down (Fig. 3b, e). In this demonstration, we employ a silicon (Si) wafer with a 1 μm thick thermal oxide (silicon dioxide, $SiO_2$) layer. After UV curing, the PDMS is peeled off, and the sample is cleaned with acetone and isopropyl alcohol to remove PPC residues on the spalled BTO film (Fig. 3c, f). While minor pitting is observed around the spalled area, likely from the interaction between the solvent and the PPC residue adhering to UV resin, the spalled region remains clean and intact. The tight bond between the film and resin preserves surface roughness during transfer.

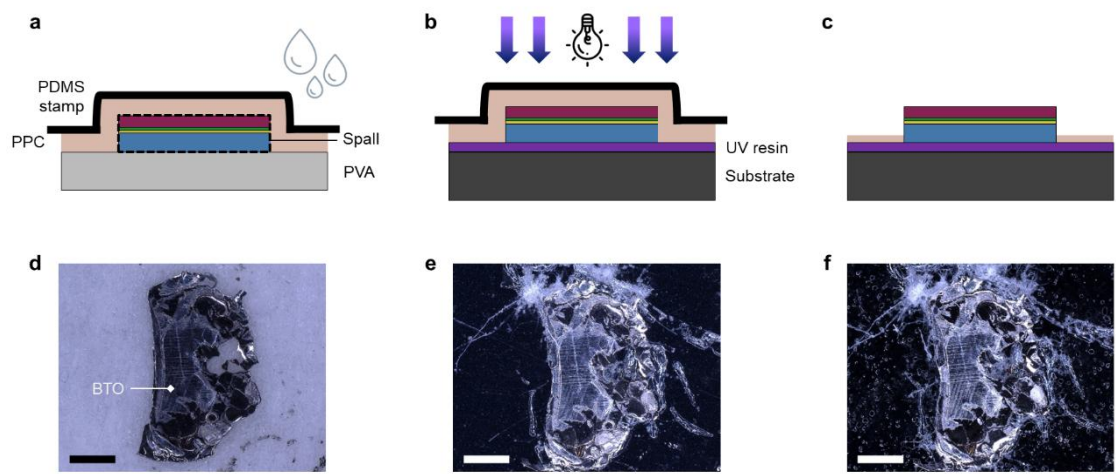

**Figure 3. Transfer process of spalled film onto substrate.** (a) The spalled heterostructure (outlined by dashed black lines), consisting of Ni (blue), Au (yellow), Ti (green), and BTO (magenta), and adhered to PVA tape (grey), is spin-coated with PPC (tan) and picked up with a PDMS stamp (black). The PVA tape is then dissolved in water. (b) The spall is picked up with PPC/PDMS and adhered to a substrate (dark grey) with UV-curable resin (purple). The resin is cured under UV light (purple arrows). (c) After curing, the sample is cleaned with acetone/IPA to remove PPC from the spalled BTO. (d)-(f) Dark field optical microscope images of (d) spalled BTO film adhered to PVA and covered with PPC and PDMS, (e) spalled film adhered to PPC/PDMS on Si/$SiO_2$ wafer with UV resin, and (f) the transferred film after cleaning in acetone/IPA. Scale bars: 1 mm.

For applications requiring Ni removal, a two-step tape transfer process can be employed: first spalling with heat-release tape, then transferring the structure to PVA tape with the spalled BTO facing down. Once the spalled heterostructure is flipped, the procedure in Fig. 3 can be followed with Ni on top. The Ni layer can then be removed using Ni etchant, though this approach relies on the resistance of the UV-curable resin to the etchant.

To evaluate the electro-optic properties of spalled BTO thin films, we characterize their Pockels coefficient, a key parameter for applications. Using the transferred film shown in Fig. 3f, we fabricate an electro-optically tunable device. The spalled film, obtained from a (001)-oriented bulk crystal is ~20 μm thick and ~0.5 x 1.5 mm$^2$ in size (Fig. 4a). The underlying Ni stressor layer acts both as back contact and reflective surface. A transparent top electrode, consisting of a 50 nm thick indium tin oxide (ITO) layer (light green dashed region in Fig. 4a), is patterned using shadow masks. The ITO layer is partially overlaid with a 5 nm Ti and 100 nm Au layer (brown dotted region in Fig. 4a) for wire bonding. This configuration allows direct characterization of the $r_{33}$ coefficient under an out-of-plane modulating field, given the c-axis orientation of the spalled BTO film.

We use the Teng-Man technique, originally developed for measuring $r_{33}$ in poled polymer films,[44,45] to determine the relative phase difference between the s- (TE) and p-polarized (TM) components of a beam reflected from the sample under a modulating voltage.[46] The experimental setup is shown in Fig. 4b, with the inset providing a schematic of the bulk BTO sample used for calibration (see Supplementary Information Part 5). The incident laser beam is linearly polarized at 45°, ensuring equal amplitude for the s- and p-polarized components. Upon reflection from the sample, the beam propagates through a Soleil-Babinet compensator (SBC), an analyzer, and into a detector. The SBC introduces a controllable retardance $\Psi_c$ through positional adjustment. The setup is used in two configurations: (1) with irises limiting the probing diameter to less than 300 μm and ensuring single-pass signal collection, and (2) with lenses in place of the irises to probe smaller areas with tens of microns diameter.

We first measure the average output intensity, expressed as

$$I_{DC} = I_0 + 2I_c \cdot \sin^2\left(\frac{\Psi_{sp}+\Psi_c}{2}\right) = C_0 + C_1 \cdot \sin^2\left(\frac{\Psi_{sp}+\Psi_c}{2}\right) \tag{1}$$

Here, $I_0$ represents the background intensity, and $I_c$ corresponds to half the maximum intensity. $\Psi_{sp}$ is the phase retardance between the s- and p-waves introduced by the sample, and $\Psi_c$ is given by $\Psi_c = C_2 \cdot x$, where $x$ is the SBC position in millimeters. By fitting the DC intensity measurements at various SBC positions to equation (1), we can extract $C_0$, $C_1$, $\Psi_{sp}$, and $C_2$.

Next, a modulating voltage $V = V_m \cdot \sin(\omega_m t)$ at frequency $\omega_m = 2\pi f_m$ is applied across the sample. This alters refractive index and optical path length, inducing an additional phase change $\delta\Psi = \left(\frac{2\pi}{\lambda}\right)(s\delta n + n\delta s)$, with $\lambda$ corresponding to the optical wavelength. The modulated intensity, $I_m$, then becomes[47]

$$I_m = \delta C_0 + \delta C_1 \cdot \sin^2\left(\frac{\Psi_{sp}+\Psi_c}{2}\right) + \frac{C_1}{2} \cdot \sin(\Psi_{sp}+\Psi_c)\delta\Psi_{sp} \tag{2}$$

The second term corresponds to reflectivity changes due to refractive index variation, while the third term reflects contributions from the phase difference change $\delta\Psi_{sp}$. Fitting $|I_m|$ at several SBC positions yields $\delta C_0$, $\delta C_1$, and $\delta\Psi_{sp}$.

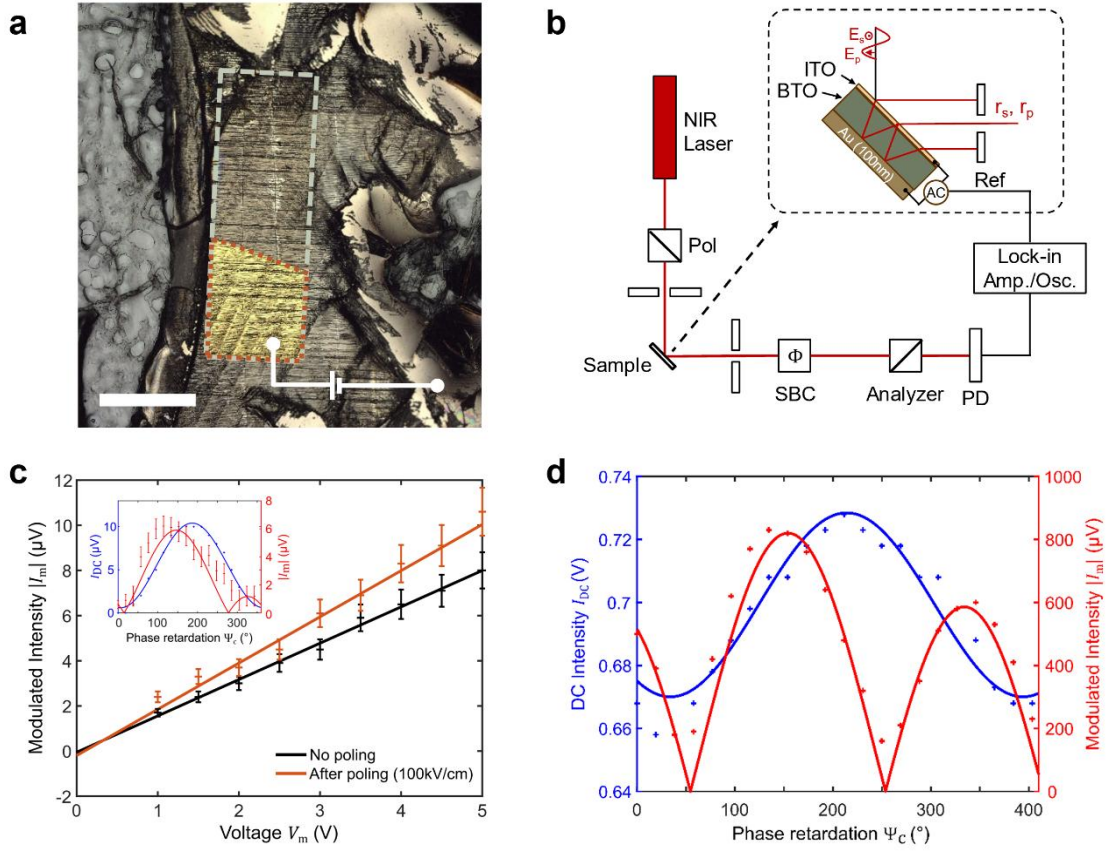

**Figure 4. Electro-optic measurement of spalled BTO thin film.** (a) Optical microscope image of BTO spall from Fig. 3f with ITO/Au top electrode. The light green dashed region indicates the area with transparent ITO for optical probing, the brown dotted region represents the area with Au on top of ITO. Wires are bonded to the Au top contact and Ni back layer, enabling probing of the electro-optic coefficients with an out-of-plane electric field. Scale bar: 500 μm. (b) Schematic of Teng-Man measurement setup. The setup was verified using bulk BTO samples, consisting of a 100 nm thick Au back electrode and a 50 nm thick ITO top electrode (inset). Configuration (1) uses irises before and after the sample. In configuration (2), each iris is replaced with a lens with a focal length of $f = 3.5$ cm. Pol: Polarizer. PD: Photodetector. Lock-in Amp: Lock-in amplifier. Osc: Oscilloscope. Ref: Reference signal. (c) Modulated intensity $|I_m|$ as a function of applied voltage $V_m$ using configuration (1) of the Teng-Man setup. Inset: DC intensity $I_{DC}$ (blue) and modulated intensity $|I_m|$ (red) as function of phase retardation $\Psi_c$ for the sample shown in (a). Crosses represent measurement points, and the lines are the fits used to extract the Pockels coefficient. Error bars correspond to standard deviations and are deduced from repeated measurements. (d) DC intensity $I_{DC}$ (blue) and modulated intensity $|I_m|$ (red) as function of phase retardation $\Psi_c$ with configuration (2) of Teng-Man setup.

The Pockels coefficients are derived using $\delta\Psi_{sp} = \Gamma_m \sin(\omega_m t)$ with

$$\Gamma_m = \left(\frac{2\pi r_{33} V_m}{\lambda}\right) \left[\frac{n_o n_e \sin^2\theta}{(n_e^2 - \sin^2\theta)^{1/2}} + \frac{r_{13}}{r_{33}} \left(\frac{n_o^3}{n_e}(n_e^2 - \sin^2\theta)^{1/2} - \frac{n_o^4}{(n_o^2 - \sin^2\theta)^{1/2}}\right)\right]^{-1} \quad (3)$$

Here, $\theta$ is the incident angle of light on the sample. When $I_m$ is measured at the half intensity point $I_c$, the modulated intensity is in its most liner region. Here, $I_m/I_c \approx \delta\Psi_{sp}$, leading to

$$r_{33} = \left(\frac{\lambda}{2\pi}\right)\left(\frac{I_m}{I_c V_m}\right)\left[\frac{n_o n_e \sin^2\theta}{(n_e^2-\sin^2\theta)^{1/2}} + \zeta\left(\frac{n_o^3}{n_e}(n_e^2-\sin^2\theta)^{1/2} - \frac{n_o^4}{(n_o^2-\sin^2\theta)^{1/2}}\right)\right]^{-1} \quad (4)$$

with $\zeta \equiv r_{13}/r_{33}$. Assuming $n_o \approx n_e \approx n$ and $\zeta = 1/10$ (from bulk BTO),[4] equation (4) simplifies to

$$r_{33} = \left(\frac{\lambda}{2\pi}\right)\left(\frac{I_m}{I_c V_m}\right) \cdot \frac{10(n^2-\sin^2\theta)^{1/2}}{9n^2\sin^2\theta} \quad (5)$$

In our measurement, light is incident at $\theta = 45°$. We validated the setup using bulk BTO and obtained a Pockels coefficient of $r_{33} = 96.4$ pm/V, consistent with literature values.[4] Further details on this measurement are provided in Supplementary Information Part 5, along with a discussion on using the Teng-Man approach to measure $r_{42}$, outlined in Supplementary Information Part 6.

Next, we characterized the spalled film shown in Fig. 4a using configuration (1) of the Teng-Man setup. The inset of Fig. 4c shows representative data for $I_{DC}$ (blue) and $|I_m|$ (red). In our measurement, we apply a modulating voltage with $V_m = 5$ V and frequency $f_m = 10$ kHz, enabling a characterization of the unclamped electro-optic coefficient.[27,48] By fitting the experimental data to equations (1) and (2) and evaluating $r_{33}$ using equation (4), we determine an electro-optic coefficient of $r_{33} = 42 \pm 3$ pm/V at $\lambda = 1500$ nm using refractive index values from Ref. [22]. To confirm that the response is governed by the electro-optic Pockels effect, we demonstrate a linear relationship between $V_m$ and $|I_m|$ in Fig. 4c (black line). The values for $|I_m|$ in Fig. 4c are recorded at $\Psi_c \sim 115°$, where the signal is maximized.

The inset of Fig. 4c shows an asymmetry in the two peaks for $|I_m|$, which is attributed to surface roughness leading to scattering, as well as partial polarization rotation due to domain switching during spalling. To reduce the impact of domain switching, we pole the spalled thin film out-of-plane by applying a constant voltage of 20 V for one hour. This procedure increases $r_{33}$ to $55 \pm 5$ pm/V. Assuming the same ratio of $r_{42}/r_{33}$ as in bulk BTO, we project a value of $r_{42} = 680$ pm/V. Although this value is lower than in bulk BTO, it exceeds values achieved using alternative bottom-up growth techniques such as sputtering or pulsed laser deposition.

To investigate the effect of domain structure on the measured electro-optic coefficient, we perform a secondary measurement using configuration (2) for probing across tens of microns in regions with fewer domains, as highlighted in Fig. 2e. Fig. 4d shows measured data for $I_{DC}$ and $|I_m|$ in blue and red, respectively. Fitting the data from a series of measurements yields an electro-optic coefficient of $r_{33} = 160 \pm 40$ pm/V. These values remain consistent up to modulation frequencies of 18 kHz, with bandwidth limitations likely arising from the RC time constant of the device under consideration (see Supplementary Information Part 7). Despite the nonideal roughness and domain structure of spalled BTO, these results suggest that bulk properties are preserved locally, in likely single-domain regions. We anticipate that future improvements to the electro-optic properties of spalled films will involve obtaining large-area thin films with low roughness or polishing thin films after spalling. Additional poling at elevated fields and temperatures may enable complete restructuring and stabilization of domains in the desired configuration.

In summary, we have developed a scalable method for fabricating high-quality single-crystal BTO thin films suitable for integration in advanced functional devices. Spalled films can be obtained with desired crystal orientations, as demonstrated for (100)- and (001)-oriented bulk crystals, with film thicknesses ranging from 100 nm to 15 μm and lateral dimensions spanning hundreds of microns to several millimeters. The film thickness is controlled by the electroplated Ni layer, which dictates the stress profile. While stress-induced domain switching introduces surface roughness during spalling, the intrinsic properties of bulk BTO are preserved. We measured an electro-optic coefficient of $r_{33} = 55 \pm 5$ pm/V in multi-domain regions and $r_{33} = 160 \pm 40$ pm/V in smaller, likely single-domain areas. Our results indicate that spalled BTO films can exceed the performance of commercially available thin-film lithium niobate, showing promise for photonic device integration. Future work will focus on improving surface quality, refining domain structures, and transferring isolated BTO membranes for broader applications.

**Acknowledgments**

This work was supported by the Air Force Office of Scientific Research Meta-Imaging MURI grant #FA9550-21-1-0312 (P.T., M.T., J.B., H.A.A.) and Caltech Space Solar Power Project (A.W.N., P.R.J.). P.T. acknowledges support from Meta Platforms, Inc., through the PhD fellowship #C-834952. J.B. acknowledges the support of the Natural Sciences and Engineering Research Council of Canada (NSERC) through the Postgraduate Scholarship – Doctoral program. We further thank Samuel K. W. Seah and Kyle Hunady for insightful discussions regarding thin film transfer.

**Author Contributions**

P.T. and H.A.A. conceived the original idea for this project. P.T. and A.W.N. performed spalls and thin film characterization, with inputs from P.R.J. P.T. developed the thin film transfer technique together with J.B.. P.T. prepared samples for electro-optic testing, M.T. built the Teng-Man setup, and P.T. and M.T. and performed the corresponding measurements. All authors contributed to writing the manuscript.

**Conflict of Interest**: The authors declare no conflict of interest.

**Data Availability Statement:** The data that support the findings of this study are available from the corresponding author upon reasonable request.

**References**

1.  McMahon, P. L. The physics of optical computing. *Nature Reviews Physics* **5**, 717–734 (2023).

2.  Hu, J. *et al.* Diffractive optical computing in free space. *Nat Commun* **15**, (2024).

3.  Miller, D. A. B. Attojoule Optoelectronics for Low-Energy Information Processing and Communications. *Journal of Lightwave Technology* **35**, 346–396 (2017).

4.  Karvounis, A., Timpu, F., Vogler-Neuling, V. V., Savo, R. & Grange, R. Barium Titanate Nanostructures and Thin Films for Photonics. *Advanced Optical Materials* vol. 8 Preprint at https://doi.org/10.1002/adom.202001249 (2020).

5.  Margalit, N. *et al.* Perspective on the future of silicon photonics and electronics. *Applied Physics Letters* vol. 118 Preprint at https://doi.org/10.1063/5.0050117 (2021).

6.  Dutt, A., Mohanty, A., Gaeta, A. L. & Lipson, M. Nonlinear and quantum photonics using integrated optical materials. *Nature Reviews Materials* vol. 9 321–346 Preprint at https://doi.org/10.1038/s41578-024-00668-z (2024).

7. Li, N. *et al.* Large-area metasurface on CMOS-compatible fabrication platform: Driving flat optics from lab to fab. *Nanophotonics* vol. 9 3071–3087 Preprint at https://doi.org/10.1515/nanoph-2020-0063 (2020).

8. Sun, J., Timurdogan, E., Yaacobi, A., Hosseini, E. S. & Watts, M. R. Large-scale nanophotonic phased array. *Nature* **493**, 195–199 (2013).

9. Sokhoyan, R., Hail, C. U., Foley, M., Grajower, M. Y. & Atwater, H. A. All-Dielectric High-Q Dynamically Tunable Transmissive Metasurfaces. *Laser Photon Rev* **18**, (2024).

10. Marris-Morini, D. *et al.* Recent progress in high-speed silicon-based optical modulators. *Proceedings of the IEEE* **97**, 1199–1214 (2009).

11. Andersen, T. I. *et al.* Beam steering at the nanosecond time scale with an atomically thin reflector. *Nat Commun* **13**, (2022).

12. Li, M., Hail, C. U., Biswas, S. & Atwater, H. A. Excitonic Beam Steering in an Active van der Waals Metasurface. *Nano Lett* **23**, 2771–2777 (2023).

13. Huang, L. *et al.* Enhanced light-matter interaction in two-dimensional transition metal dichalcogenides. *Reports on Progress in Physics* vol. 85 Preprint at https://doi.org/10.1088/1361-6633/ac45f9 (2022).

14. Zhu, D. *et al.* Integrated photonics on thin-film lithium niobate. *Adv Opt Photonics* **13**, 242 (2021).

15. Boes, A. *et al.* Lithium niobate photonics: Unlocking the electromagnetic spectrum. *Science* vol. 379 Preprint at https://doi.org/10.1126/science.abj4396 (2023).

16. Thomaschewski, M., Zenin, V. A., Wolff, C. & Bozhevolnyi, S. I. Plasmonic monolithic lithium niobate directional coupler switches. *Nat Commun* **11**, (2020).

17. Wang, C., Zhang, M., Stern, B., Lipson, M. & Lončar, M. Nanophotonic lithium niobate electro-optic modulators. *Opt Express* **26**, 1547 (2018).

18. Qi, Y. & Li, Y. Integrated lithium niobate photonics. *Nanophotonics* vol. 9 1287–1320 Preprint at https://doi.org/10.1515/nanoph-2020-0013 (2020).

19. Adediji, Y. B., Adeyinka, A. M., Yahya, D. I. & Mbelu, O. V. A review of energy storage applications of lead-free BaTiO3-based dielectric ceramic capacitors. *Energy, Ecology and Environment* vol. 8 401–419 Preprint at https://doi.org/10.1007/s40974-023-00286-5 (2023).

20. Jiang, B. *et al.* Barium titanate at the nanoscale: Controlled synthesis and dielectric and ferroelectric properties. *Chemical Society Reviews* vol. 48 1194–1228 Preprint at https://doi.org/10.1039/c8cs00583d (2019).

21. Japze, H. *Elastic and Piezoelectric Coefficients of Single-Crystal Barium Titanate*. (1958).

22. Abel, S. *et al.* A hybrid barium titanate-silicon photonics platform for ultraefficient electro-optic tuning. *Journal of Lightwave Technology* **34**, 1688–1693 (2016).

23. Girouard, P. *et al.* Modulator with 40-GHz Modulation Utilizing BaTiO3 Photonic Crystal Waveguides. *IEEE J Quantum Electron* **53**, (2017).

24. Sun, D. *et al.* Theoretical Feasibility Demonstration for over 100 GHz Electro-Optic Modulators with c-Axis Grown BaTiO 3 Crystal Thin-Films. *Journal of Lightwave Technology* **33**, 1937–1947 (2015).

25. Jin, T. & Lin, P. T. Efficient Mid-Infrared Electro-Optical Waveguide Modulators Using Ferroelectric Barium Titanate. *IEEE Journal of Selected Topics in Quantum Electronics* **26**, (2020).

26. Palik, E. D. . & Ghosh, Gorachand. *Handbook of Optical Constants of Solids Five-Volume Set*. (Academic Press, 1998).

27. Fontana, M. D., Laabidi, K., Jannot, B., Maglione, M. & Jullien, P. Relationship between electro-optic, vibrational and dielectric properties in BaTiO3. *Solid State Commun* **92**, 827–830 (1994).

28. Abel, S. *et al.* Large Pockels effect in micro- and nanostructured barium titanate integrated on silicon. *Nat Mater* **18**, 42–47 (2019).

29. Winiger, J. *et al.* PLD Epitaxial Thin-Film BaTiO3 on MgO – Dielectric and Electro-Optic Properties. *Adv Mater Interfaces* **11**, (2024).

30. Dong, Z. *et al.* Monolithic Barium Titanate Modulators on Silicon-on-Insulator Substrates. *ACS Photonics* **10**, 4367–4376 (2023).

31. Kormondy, K. J. *et al.* Microstructure and ferroelectricity of BaTiO3 thin films on Si for integrated photonics. *Nanotechnology* **28**, (2017).

32. Weigand, H. C. *et al.* Nanoimprinting Solution-Derived Barium Titanate for Electro-Optic Metasurfaces. *Nano Lett* **24**, 5536–5542 (2024).

33. Thureja, P. *et al.* Toward a universal metasurface for optical imaging, communication, and computation. *Nanophotonics* **11**, 3745–3768 (2022).

34. Wessels, B. W. Ferroelectric epitaxial thin films for integrated optics. *Annual Review of Materials Research* vol. 37 659–679 Preprint at https://doi.org/10.1146/annurev.matsci.37.052506.084226 (2007).

35. Chen, J. & Packard, C. E. Controlled spalling-based mechanical substrate exfoliation for III-V solar cells: A review. *Solar Energy Materials and Solar Cells* vol. 225 Preprint at https://doi.org/10.1016/j.solmat.2021.111018 (2021).

36. Hutchinson, J. W. . *Advances in Applied Mechanics Vol 29*. (Academic Press, 1992).

37. Lee, Y. H., Kim, J. & Oh, J. Wafer-Scale Ultrathin, Single-Crystal Si and GaAs Photocathodes for Photoelectrochemical Hydrogen Production. *ACS Appl Mater Interfaces* **10**, 33230–33237 (2018).

38. Bedell, S. W. *et al.* Kerf-less removal of Si, Ge, and III-V layers by controlled spalling to enable low-cost PV technologies. *IEEE J Photovolt* **2**, 141–147 (2012).

39. Bedell, S. W. *et al.* Layer transfer by controlled spalling. *J Phys D Appl Phys* **46**, (2013).

40. Crouse, D. Controlled Spalling in (100)-Oriented Germanium by Electroplating. (Colorado School of Mines, 2017).

41. Sweet, C. A. Spalling Fracture Behavior in (100) Gallium Arsenide. (Colorado School of Mines, 2016).

42. Prasad, V. C. S. & Subbarao, E. C. Deformation and stress relaxation of single crystal BaTiO3. *Ferroelectrics* **15**, 143148 (1977).

43. Meschke, F., Koll~eck, A. & Schneider, G. A. *R-Curve Behaviour of BaTi03 Due to Stress-Induced Ferroelastic Domain Switching*. *Journal of the European Ceramic Society* vol. 17 (1997).

44. Teng, C. C. & Man, H. T. Simple reflection technique for measuring the electro-optic coefficient of poled polymers. *Appl Phys Lett* **56**, 1734–1736 (1990).

45. Schildkraut, J. S. *Determination of the Eiectrooptic Coefficient of a Poled Polymer Film*. vol. 29 (1990).

46. Shuto, Y. & Amano, M. Reflection measurement technique of electro-optic coefficients in lithium niobate crystals and poled polymer films. *J Appl Phys* **77**, 4632–4638 (1995).

47. Inoue, A., Inoue, S. I. & Yokoyama, S. Enhanced electro-optic response of a poled polymer in a reflective microcavity. *Opt Commun* **283**, 2935–2938 (2010).

48. Chelladurai, D. *et al.* Barium Titanate and Lithium Niobate Permittivity and Pockels Coefficients from MHz to Sub-THz Frequencies. *arXiv:2407.03443 [physics.optics]*.

Supplementary Information for

# Spalled barium titanate single crystal thin films for functional device applications

*Prachi Thureja,[1] Andrew W. Nyholm,[1] Martin Thomaschewski,[1] Phillip R. Jahelka,[1] Julie Belleville,[1] and Harry A. Atwater[1,*]*

[1]Thomas J. Watson Laboratories of Applied Physics, California Institute of Technology, Pasadena, California 91125, USA

*Corresponding author: haa@caltech.edu

## 1. XRD scans for bulk substrates and spalled film

Commercially bought BTO substrates[1] were produced *via* top-seeded solution growth and were single-crystalline with multiple domains. We confirmed the single-crystallinity of bulk substrates and spalled films through θ-2θ x-ray diffraction (XRD) scans. As shown in Fig. S1 illustrate, the prominent peaks for both the bulk substrate and spalled thin film correspond to the (001) and (100) orientations, along with their higher-order diffractions. These peaks arise due to the multi-domain nature of the substrates and films, combined with the millimeter-scale x-ray probing spot. To probe the domain structure at a finer scale, we employed electron backscatter diffraction (EBSD), as discussed in the manuscript. Similar results were obtained for (100)-oriented crystals.

Figure S1b shows the θ-2θ scan for a thin film spalled from a (001)-oriented single-crystal. The reduced intensity observed for the thin film is primarily attributed to its surface roughness and reduced thickness. In addition to the signal from BTO, we also measure a minimal contribution from the underlying Ni layer due to the size and position of the x-ray probing spot relative to the spalled film area.

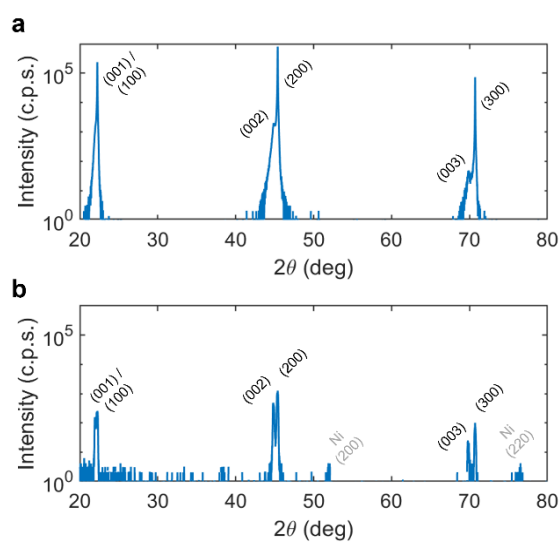

**Figure S1. XRD data for (001)-oriented bulk substrate and spalled thin film.** θ-2θ scan for (a) bare bulk substrate and (b) spalled thin film on a Ti/Au seed layer and electroplated Ni. BTO diffraction

orders corresponding to various crystal orientations are labelled in both figures, Ni diffraction orders are indicated in grey in (b).

## 2. Reusability of single-crystal BTO substrates through repolishing

One major advantage of spalling compared to alternative growth techniques is the ability to reuse bulk substrates by introducing a polishing step between spalls. Considering that the substrates are typically 0.5-1 mm thick, and each iteration removes less than 50 μm of material thickness, the effective cost of a spall can be lowered up to 20-fold, bringing the expense down to tens of dollars per square centimeter.

In this study, various polishing procedures were tested, and the RMS roughness $r$ of each was compared to that of as-bought substrates. Figure S2a shows an AFM scan of an as-bought substrate with $r = 5.9$ nm. Hand polishing of BTO substrates yields planarized surfaces but results in a significantly higher RMS roughness of $r = 66.4$ nm. While spalls can still be performed with level of roughness, it is insufficient for many applications.

To address this, we explored vibration polishing using a slurry of silica ($SiO_2$) or alumina ($Al_2O_3$) nanoparticles, which significantly reduce surface roughness. Figures S2b and c show AFM scans of the BTO substrate after vibration polishing with 60 nm-sized $SiO_2$ particles for 7 hours and 50 nm-sized $Al_2O_3$ particles for 3 hours, respectively. The resulting RMS roughness values were $r = 27.2$ nm and $r = 9.4$ nm. The stronger reduction in surface roughness achieved with $Al_2O_3$ particles is attributed to their higher material hardness relative to BTO.

It is important to note that while vibration polishing does not planarize the substrate, it effectively restores surface roughness to near-initial levels, enabling large-scale spalls suitable for device integration, as shown in Fig. 1b of the manuscript.

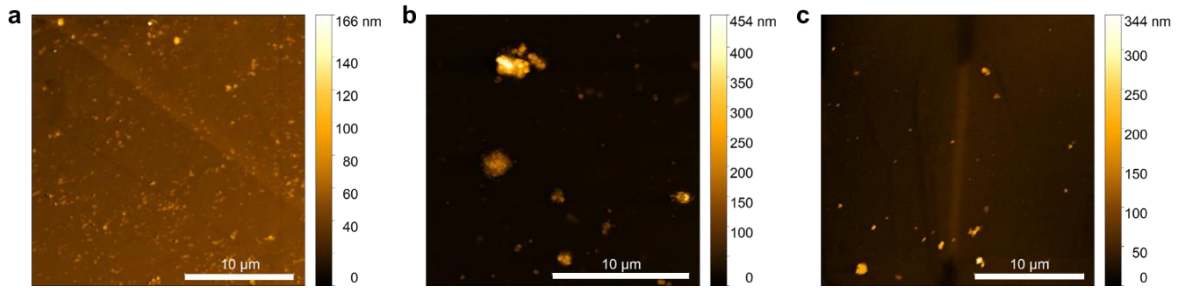

**Figure S2. AFM scans of BTO substrate as-bought *vs* polished after spalling.** (a) As-bought BTO substrate, (b) as-bought BTO substrate, vibration polished with 60 nm-sized $SiO_2$ particles for 7 hours, (c) BTO substrate post-spall, vibration polished with 50 nm-sized $Al_2O_3$ particles for 3 hours. Scale bars: 10 μm.

## 3. Thickness and roughness variability for (100) oriented spalls

In the manuscript, we described methods to control the thickness – and consequently the roughness – of BTO spalls obtained from (001)-oriented bulk crystals (*c*-axis). Here, we present the results for a similar study conducted on (100)-oriented bulk crystals (*a*-axis). Figure S3a illustrates spalled BTO thin films obtained by electroplating three different Ni thicknesses on the same bulk crystal. These three regions were achieved by electroplating two partially overlapping Ni layers, following a method similar to the one described in the manuscript. Specifically, the three stress states were created by first electroplating approximately 2 μm of Ni, followed by a second layer with a thickness of about 600-800 nm.

Figure S3a shows the spall performed using PVA tape. The highest stress state is located in the central region (blue box), while the intermediate and low-stress regions are on the left (orange box) and right (green box), respectively. Due to the small difference in the Ni stressor layer thickness between the intermediate and high-stress regimes, only minimal differences in roughness are observed in the spalled films from these two regions. The spall depth $t$, measured *via* profilometry, ranges around 6-11 μm in most regions of the intermediate stress regime, while values between $t = 10\text{-}12$ μm are obtained for the high-stress regime.

Notably, finer features on the spalled films, such as horizontally or vertically oriented domain lines, extend across both the intermediate and high-stress regions, as seen in the blue box in image S3b. These features correspond to those found on the bulk substrate used for spalling, suggesting that the surface quality and domain structure of the bulk substrates are critical factors influencing the spalled thin films (see Supplementary Information Part 4 for further discussion). While the surface quality is comparable between most parts of the high and intermediate-stress spalls, the second panel (orange box) in Fig. S3b reveals that small regions exhibiting thin-film interference can also be observed in the (100)-oriented films under intermediate stress conditions.

These thin film interference regions can be achieved more controllably by electroplating a thinner Ni layer, as shown in the low-stress regime. Because the Ni stressor thickness used in this case was smaller than that used for the (001)-oriented substrate in the manuscript, we obtain smaller spalled areas here. Nevertheless, we can identify thin film areas spanning up to tens of microns laterally, identified through the thin-film interference effect, as shown in the optical microscope image (green box, Fig. S3b). Due to the inherently lower surface quality of the (100)-oriented bulk BTO substrate, we observe some film breakage within the thin-film regions.

Overall, this lower substrate surface quality contributes to increased surface roughness in spalled films obtained from (100)-oriented crystals compared to those from (001)-oriented substrates. Figure S3c provides AFM scans of three different regions within the corresponding panels in Fig. S3b. The RMS roughness was evaluated over the white dashed regions in Fig. S3c and corresponds to approximately 235 nm in the high-stress state (top), 56 nm in the intermediate-stress regime (middle), and 47 nm for the low-stress regime (bottom). To find the correlation between film thickness and roughness across these regions, additional scans were performed in different areas, and the results are shown in Fig. 2d of the manuscript.

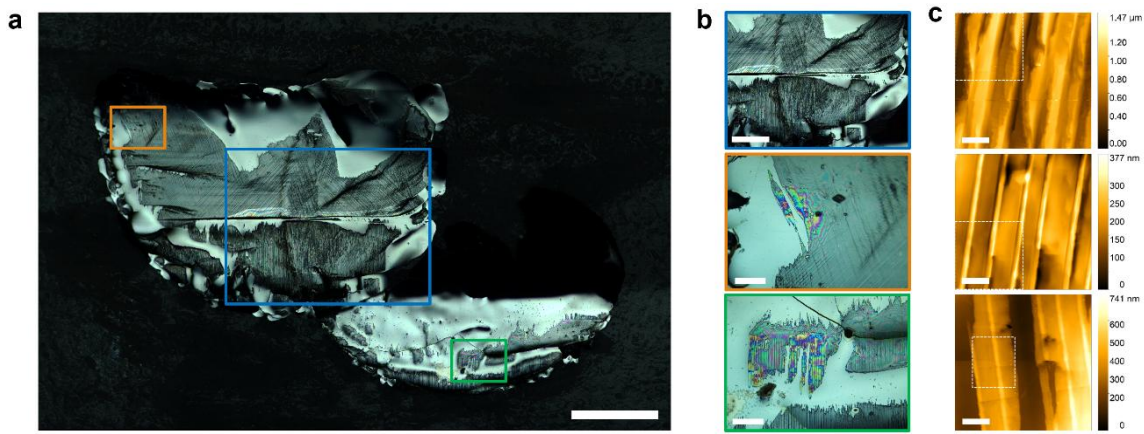

**Figure S3. Spalled BTO films from (100) bulk substrates with varying thicknesses.** (a) Scale bar: 1 mm. (b) Scale bars (from top to bottom): 500 μm, 100 μm, 100 μm. (c) AFM images of spalled (100) thin films within the areas shown in (b). Dashed white region in each image corresponds to a region over which the RMS roughness was evaluated. Scale bars: 10 μm.

## 4. Surface quality of (100) and (001)-oriented bulk crystals

Commercially bought single crystal BTO substrates[1] are produced using top-seeded solution growth (TSSG). This process involves growing boules from molten solutions at elevated temperatures. During this process, the controlled cooling and crystallization of the molten material enable the formation of high-quality single crystals.[2] After the boule is formed, sections of it are cut and polished to create substrates. These substrates are typically multidomain in nature, with distinct regions corresponding to different orientations of the polar axis (spontaneous polarization).

This multidomain structure forms as BTO undergoes a structural phase transition from the cubic to tetragonal phase upon cooling below its Curie temperature. This transition generates spontaneous polarization, with the formation of domains being influenced by various factors such as the cooling rate, internal and external stresses, and thermal gradients during TSSG.[3] These conditions lead to the formation of 90° and 180° domain walls, which separate regions with differing polarization orientations. Further details on the domain structure of BTO can be found in Refs. [3–7].

We note significant differences in the bulk substrate quality between (100)- and (001)-oriented crystals, as illustrated in Fig. S4. The (100)-oriented substrates feature prominent 90° domain patterns and finer surface ridges (domain bundles). In contrast, (001)-oriented substrates exhibit a more uniform appearance with fewer domain boundaries. This discrepancy likely arises from the anisotropic strain in the in-plane directions of $a$-axis crystals during cooling, which promotes the formation of additional domain walls. The higher domain density in (100)-oriented substrates results in increased surface roughness. The elevated roughness carries over to spalled thin films, leading to a more irregular topography. By contrast, the smoother surface of (001)-oriented substrates yields spalled films with more controllable and uniform surface properties.

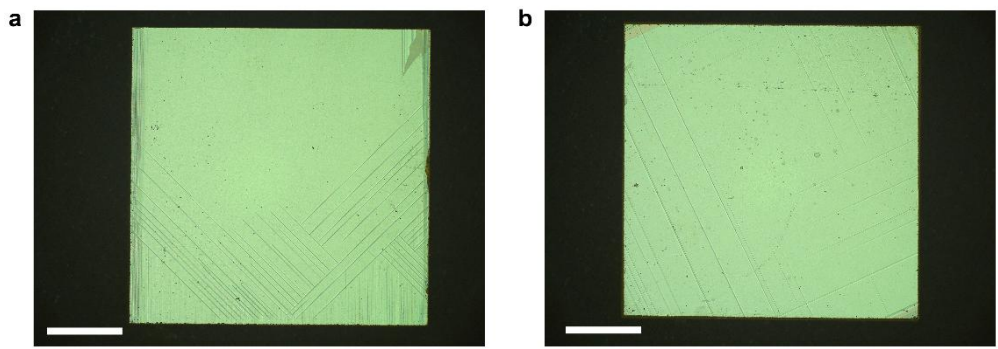

**Figure S4. Single crystal BTO substrates with domains.** Optical microscope image of (a) (100)-oriented single crystal substrate with domains, and (b) (001)-oriented single crystal substrate with domains. Scale bars: 2.5 mm.

## 5. Electro-optic characterization of bulk BTO substrate

To calibrate the Teng-Man measurement technique and obtain reference values for the electro-optic coefficients of bulk BTO substrates used for spalling, we pattern (100)-oriented bulk substrates ($a$-axis) with top and bottom electrodes. The bottom electrode consists of a 5 nm Ti adhesion layer followed by a 100 nm Au layer and is contacted from the edge of the substrate using silver paste. For the top electrodes, we pattern three 50 nm thick ITO contacts, as shown in Fig. S5a. These top contacts are designed to enable probing of the electro-optic coefficients with both in-plane and out-of-plane electric fields. Each ITO top electrode also includes an Au bond pad (5 nm Ti / 80 nm Au) for electrical addressing.

The choice of crystal orientation for bulk testing allows us to potentially probe both $r_{33}$ and $r_{42}$ coefficients (see Supplementary Information Part 6 for further discussion), unlike (001)-oriented substrates (c-axis), which primarily facilitate $r_{33}$ measurements. For electro-optic testing of spalled films discussed in the manuscript, we use films with a primary (001)-orientation. This selection is motivated by the improved surface roughness of (001)-oriented films and the availability of mm-scale films, which enable easy patterning of top contacts using shadow masks. This approach eliminates the need for more involved fabrication methods, such as photolithography or electron beam lithography steps, which would be required to pattern multiple smaller top contacts on (100)-oriented spalled films (as shown in Fig. S5a for bulk substrates).

While the *a*-axis orientation allows measurement of additional electro-optic coefficients, it introduces in-plane birefringence. This birefringence causes the $E_x$ and $E_y$ components of the +45° linearly polarized incident plane wave to experience different refractive indices, resulting in a rotation of the polarization of the reflected beam. To account for this effect, the Teng-Man setup for *a*-axis samples requires the polarizer to be oriented at +45°, the SBC at +45° and the analyzer at approximately 0°. In contrast, *c*-axis films conventionally use a configuration with the polarizer, SBC and analyzer oriented at +45°, 90° and -45°, respectively. The measurement configurations are selected based on the expected shape of the $I_{DC}$ and $|I_m|$ curves, with $I_{DC}$ resulting in a single peak around a phase retardation of 180° and $|I_m|$ resulting in two maxima at the quadrature points of the DC curve. We refer to the first configuration as *a*-axis configuration while the second is referred to as *c*-axis configuration. All measurements are performed at a wavelength of $\lambda = 1520$ nm.

In a first step, we probe the electro-optic coefficient on top of the ITO electrode area with an out-of-plane modulating field and the Teng-Man setup in the *a*-axis configuration. Initial measurements yield an electro-optic coefficient of 4.1 pm/V. We then pole the bulk BTO substrate out-of-plane with a voltage of 200 V across a thickness of 0.5 mm (poling field of 4 kV/cm) for 1 hour. After poling, the Teng-Man setup is switched to the *c*-axis configuration to obtain clean measurements, yielding an electro-optic coefficient of 11.4 pm/V. We attribute this increase in the electro-optic coefficient to a reorientation of the domains from *a*-axis to *c*-axis in the probing region. However, due to a preferential (100) orientation of the substrate and a relatively small poling field, only partial reorientation occurs, resulting in an electro-optic coefficient smaller than the literature value for $r_{33}$.

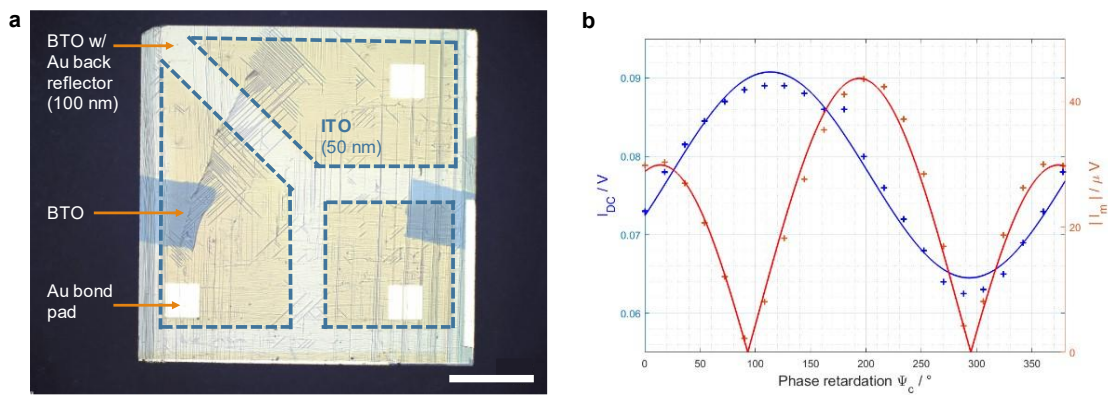

**Figure S5. Electro-optic characterization of (100) bulk BTO substrate.** (a) 0.5 mm thick (100)-oriented bulk BTO substrate with 100 nm thick Au back reflector and top electrodes consisting of 50 nm thick ITO (area inside grey dashed lines) and 1 mm² sized bond pads consisting of 5 nm Ti and 100

nm Au. Scale bar: 2.5 mm. (b) DC intensity $I_{DC}$ (blue) and modulated intensity $|I_m|$ (red) obtained for bulk substrate shown in (a).

To probe $r_{33}$ more accurately with a (100)-oriented bulk substrate, we move the probing spot to an area between two ITO top electrodes in the bottom half of the substrate shown in Fig. S5a. This position allows application of an in-plane modulating field in a preferentially $a$-axis oriented region. We perform the measurements using an $a$-axis configuration of the Teng-Man setup and obtain an electro-optic coefficient of 22.3 pm/V. We then apply a voltage of 200 V between the two in-plane electrodes spaced 1 mm apart (poling field of 2 kV/cm), leading to a drastic increase in the measured electro-optic coefficient, which reaches 96.4 pm/V. This value is in good agreement with literature values for bulk BTO ($r_{33,lit}$ = 105 ± 10 pm/V)[8]. The corresponding measurements for $I_{DC}$ and $|I_m|$ are shown in Fig. S5b, indicating clear peaks in the modulated intensity curve (red) at the quadrature points of the DC curve (blue).

## 6. Measurement procedure for characterization of $r_{42}$

In the preceding discussion, we explored how the Teng-Man measurement setup can be configured to allow precise characterization of the $r_{33}$ coefficient. We found that our measurements approach literature values for the Pockels coefficients of BTO when the sample is poled in the preferred crystal orientation (out-of-plane poling field for $c$-axis and in-plane poling field for $a$-axis). So far, we focused on an electro-optic characterization that relies on the same electrodes for both poling and probing, resulting in parallel poling and modulating fields. However, the in-plane birefringence of $a$-axis BTO substrates provides an opportunity to additionally probe the $r_{42}$ coefficient with the Teng-Man setup.

To achieve this, an $a$-axis substrate or spalled film must first be poled using an in-plane field, as described in Supplementary Information Part 5. Following this step, instead of applying a modulating field that is parallel to the poling field, electro-optic coefficients would be probed at an angle to the poling field. This requires patterning individual probing electrodes that are integrated within the gap between the poling electrodes. These electrodes can be oriented at various angles to the poling field to enable angle-dependent characterization of the electro-optic coefficients. With this configuration, it becomes possible to probe a linear combination of $r_{33}$ and $r_{42}$. This approach is analogous to phase modulator designs utilizing an $a$-axis oriented BTO film with lateral electrodes positioned at approximately 45°, which maximize the effective electro-optic coefficient.[9]

## 7. Bandwidth of modulation signal

The Pockels tensor generally includes contributions from three components: the piezoelectric, electronic, and ionic effects.[10] At low frequencies, from DC up to approximately 1 MHz, the photoelastic response due to mechanical lattice deformations follows the modulation signal, with lattice vibrations related to acoustic phonons also contributing to the modulation.[9] Between 1 to 100 MHz, a crystal is considered to be 'clamped' because mechanical deformations are too slow to follow the modulation signal. In this regime, nonlinearities related to optical phonons dominate and are referred to as ionic contributions. Above a few THz, ionic contributions become negligible, and the electro-optic effect is primarily driven by electronic resonances with characteristic frequencies in the PHz regime. Given this frequency dependence, the Pockels tensor is typically grouped into the unclamped tensor (with contributions from all three components, up to few MHz) and the clamped tensor (higher frequency contributions).

In this work, the electro-optic Pockels coefficients were characterized using the Teng-Man setup at a frequency of 10 kHz. Figure S6 illustrates the frequency dependence of the modulated signal up to 1 MHz, showing a -3 dB cutoff frequency ($f_{-3dB}$) at approximately 18 kHz. The decrease in modulation intensity is attributed to the RC time constant of the device, rather than a transition from unclamped to clamped Pockels coefficients. The circuit model for the spalled BTO film measured for electro-optic characterization is shown in the inset of Fig. S6.

In the ideal case, $f_{-3dB}$ can be approximated using the resistance of the ITO layer ($\rho_{ITO} \sim 10^{-4}$ Ω·cm), the relative permittivity of BTO along the $c$-axis ($\epsilon_{BTO,c} \sim 135$), the dimensions of the ITO top electrode ($w_{ITO}$ = 0.5 mm, $L_{ITO}$ = 1 mm), as well as the film thicknesses ($t_{ITO}$ = 50 nm, $t_{BTO}$ = 20 μm). However, the surface roughness of the spalled BTO film will introduce deviations from these values. Based on the results shown in Fig. 2d of the manuscript, we expect greater than 200 nm surface roughness for a 20 μm thick spalled BTO film. Since the ITO thickness is much smaller than the RMS roughness of the BTO film, current flow paths will be disrupted, increasing ITO resistivity as described by the Fuchs-Sondheimer model.[11,12] Assuming an increase in ITO resistivity by two orders of magnitude ($\rho_{ITO,rough} \sim 10^{-2}$ Ω·cm), the theoretical modulation bandwidth of the spalled film can be calculated as

$$f_{-3dB} = \frac{1}{2\pi RC} = \frac{1}{2\pi} \times \frac{w_{ITO} \cdot t_{ITO}}{\rho_{ITO,rough} \cdot L_{ITO}} \times \frac{t_{BTO}}{\epsilon_0 \cdot \epsilon_{BTO} \cdot L_{ITO} \cdot w_{ITO}} \sim 592 \text{ kHz} \qquad (S1)$$

Here, the conductivities of Ni and Au are neglected. Our calculations indicate that the upper limit of the frequency bandwidth for the current device is in the range of hundreds of kHz. Empirical data shows an even lower RC time constant for the measured device, which could be attributed to an even lower ITO resistivity and increased relative permittivity of BTO due to local variations in the domain structure. To accurately capture the transition from unclamped to clamped electro-optic coefficients, alternative device geometries will need to be explored.

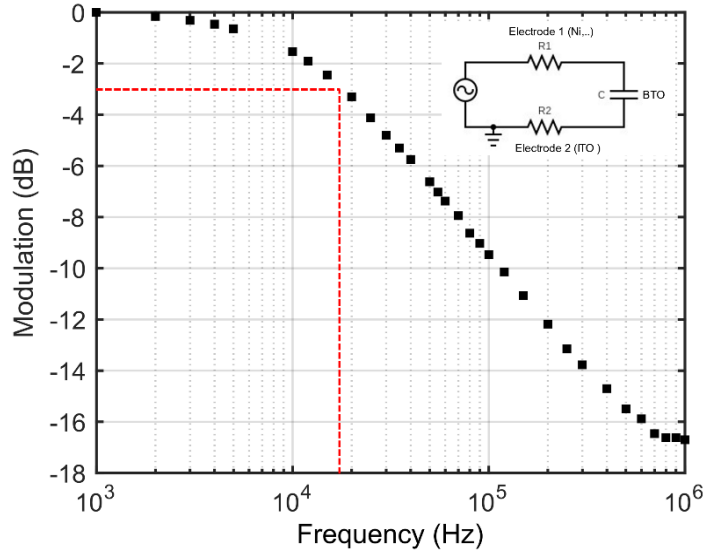

**Figure S6.** Normalized modulation intensity as a function of modulation frequency. The red dashed line indicates the -3 dB cutoff frequency at approximately 18 kHz. The inset illustrates the equivalent circuit of the device, with R1 and R2 representing the resistances of the top and bottom electrodes, and C representing the capacitance of the BTO layer.

# References

1. MTI Corp. BaTiO3 - Barium Titanate Crystal.

2. Garrett, M. H. & Mnushkina, I. Techniques for top-seeded solution growth of BaTiO3. *J Cryst Growth* **166**, 550–557 (1996).

3. Garrett, M. H., Changt, J. Y., Jenssen, H. P. & Warde, C. A method for poling barium titanate, BaTiO3. *Ferroelectrics* **120**, 167–173 (1991).

4. Bednyakov, P. S., Sluka, T., Tagantsev, A. K., Damjanovic, D. & Setter, N. Formation of charged ferroelectric domain walls with controlled periodicity. *Sci Rep* **5**, (2015).

5. Howell, J. A., Vaudin, M. D., Friedman, L. H. & Cook, R. F. Stress and strain mapping of micro-domain bundles in barium titanate using electron backscatter diffraction. *J Mater Sci* **52**, 12608–12623 (2017).

6. Howell, J. A., Vaudin, M. D., Friedman, L. H. & Cook, R. F. Lamellar and bundled domain rotations in barium titanate. *J Mater Sci* **54**, 116–129 (2019).

7. Bednyakov, P. S., Rafalovskyi, I. & Hlinka, J. Fragmented charged domain wall below the tetragonal-orthorhombic phase transition in BaTiO3. *arXiv:2410.14476v1 [cond-mat.mtrl-sci]* (2024).

8. Karvounis, A., Timpu, F., Vogler-Neuling, V. V., Savo, R. & Grange, R. Barium Titanate Nanostructures and Thin Films for Photonics. *Advanced Optical Materials* vol. 8 Preprint at https://doi.org/10.1002/adom.202001249 (2020).

9. Chelladurai, D. *et al.* Barium Titanate and Lithium Niobate Permittivity and Pockels Coefficients from MHz to Sub-THz Frequencies. *arXiv:2407.03443 [physics.optics]*.

10. Fontana, M. D., Laabidi, K., Jannot, B., Maglione, M. & Jullien, P. Relationship between electro-optic, vibrational and dielectric properties in BaTiO3. *Solid State Commun* **92**, 827–830 (1994).

11. Fuchs, K. The conductivity of thin metallic films according to the electron theory of metals. *Mathematical Proceedings of the Cambridge Philosophical Society* **34**, 100–108 (1938).

12. Sondheimer, E. H. The mean free path of electrons in metals. *Adv Phys* **50**, 499–537 (2001).